\documentclass{IEEEtran}

\usepackage[export]{adjustbox}
\usepackage{booktabs}
\usepackage{float}
\usepackage{flushend}
\usepackage{graphicx}
\usepackage{ifpdf}
\usepackage[utf8]{inputenc}
\usepackage{listings}
\usepackage{multirow}
\usepackage[caption=false,font=footnotesize]{subfig}
\usepackage{url}
\usepackage{tabularx}
\usepackage{textcomp}
\usepackage{tikz}
\usepackage{xcolor}
\usepackage{xspace}
\usepackage{cite}
\usepackage{comment}
\usepackage[nameinlink]{cleveref}

\Crefname{figure}{Fig.}{Figs.}
\newif\ifdraft{}

\ifdraft{}
  \newcommand{\mtnote}[1]{ {\textcolor{orange} { ***matteo: #1 }}}
  \newcommand{\jhanote}[1]{ {\textcolor{red} { ***shantenu: #1 }}}
  \newcommand{\miknote}[1]{ {\textcolor{brown} { ***mikhail: #1 }}}
  \newcommand{\amnote}[1]{ {\textcolor{blue} { ***andre: #1 }}}
  \newcommand{\ooknote}[1]{ {\textcolor{blue} { ***ozgur: #1 }}}
  \newcommand{\twnote}[1]{ {\textcolor{blue} { ***tianle: #1 }}}
  \newcommand{\generalnote}[1]{ {\textcolor{gray} { *note: #1 }}}

\else
  \newcommand{\mtnote}[1]{}
  \newcommand{\jhanote}[1]{}
  \newcommand{\miknote}[1]{}
  \newcommand{\amnote}[1]{}
  \newcommand{\ooknote}[1]{}
  \newcommand{\twnote}[1]{}
  \newcommand{\generalnote}[1]{}
\fi

\newcommand{\UP}{\vspace*{-1.0em}}
\newcommand{\up}{\vspace*{-0.5em}}
\newcommand{\T}[1]{\texttt{#1}\xspace}

\ifpdf{}
  \DeclareGraphicsExtensions{.pdf, .jpg, .tif}
\else
  \DeclareGraphicsExtensions{.eps, .jpg, .ps}
\fi

\newlength\myheight
\newcommand*\ccircled[1]{\settowidth{\myheight}{#1}%
    \raisebox{-.1\myheight}{\tikz[baseline=(char.base)]{%
        \node[shape=circle,draw,minimum size=\myheight*\myheight*.4,inner sep=1pt](char){#1};}}}

\begin{document}

\title{Scalable Runtime Architecture for Data-driven, Hybrid HPC and ML Workflow Applications}

\author{Andre Merzky$^{*1}$, Mikhail Titov$^{*2}$, Matteo Turilli$^{*1,2}$, Ozgur Kilic$^{2}$, Tianle Wang$^{2}$, Shantenu Jha$^{1,3,4}$\\
   \small{\emph{$^{1}$ Rutgers, the State University of New Jersey, Piscataway, NJ 08854, USA}}\\
   \small{\emph{$^{2}$ Brookhaven National Laboratory, Upton, NY 11973, USA}}\\
   \small{\emph{$^{3}$ Princeton Plasma Physics Laboratory, Princeton, NJ, USA}}\\
   \small{\emph{$^{4}$ Princeton University, Princeton, NJ, USA}}\\
   \small{$^{*}$ Contributed Equally}

}

\maketitle

\begin{abstract}
Hybrid workflows combining traditional HPC and novel ML methodologies are transforming scientific computing. This paper presents the architecture and implementation of a scalable runtime system that extends RADICAL-Pilot with service-based execution to support AI-out-HPC workflows. Our runtime system enables distributed ML capabilities, efficient resource management, and seamless HPC/ML coupling across local and remote platforms. Preliminary experimental results show that our approach manages concurrent execution of ML models across local and remote HPC/cloud resources with minimal architectural overheads. This lays the foundation for prototyping three representative data-driven workflow applications and executing them at scale on leadership-class HPC platforms.
\end{abstract}

\section{Introduction}\label{sec:intro}

For many scientific domains, integrating high-performance computing (HPC) and machine learning (ML) methodologies is starting to deliver dramatic performance gains and novel scientific insight~\cite{jha2022ai}. Following the ``learning everywhere'' paradigm~\cite{fox2019taxonomy}, integrating AI/ML methods in traditional high-performance and high-throughput calculations promises significant scientific insight~\cite{brewer2024ai}. In simple terms, AI methods can integrate ``inside'' or ``outside'' the HPC applications. Replacing parts or the entire application with a surrogate is the canonical example of AI-in-HPC. Enhancing the HPC workflow with AI methods (e.g., steering the workflow, coupling to digital twins, etc.) represents AI-out-HPC scenarios.

Whereas the new class of applications requires novel capabilities across the entire hardware and software stack, our focus in this paper will be on AI-out-HPC applications. Specifically, traditional runtime and middleware capabilities need to be extended to accommodate the novel requirements of ML technologies. On the one hand, we need to retain runtime capabilities that allow us to efficiently and effectively execute heterogeneous compute tasks at scale, whether they require MPI, CPUs, or GPUs and are implemented as executables or functions. On the other hand, we need to devise new abstractions that enable the integration of ML technologies into the runtime capabilities. Such integration must support the various execution scenarios the diverse AI-out-HPC applications require, accounting for the different ML drivers and a continuum of local and remote computing resources.

The integration of AI and HPC into AI-HPC hybrid workflows (hereafter hybrid workflows) is often achieved by using ad-hoc scripts that orchestrate the execution of specific ML and/or HPC tasks on particular resources or by using specialized systems that enable the execution of ML tasks on HPC platforms, but often with suboptimal performance and/or flexibility~\cite{ejarque2022enabling,brewer2024ai}.
These local solutions are limited in scope, not interoperable, and do not provide a common framework for designing and implementing hybrid workflows. Further, single-point solutions do not allow for devising best practices and general architectural patterns, limiting understanding of a rapidly evolving landscape.

In this paper, we present the preliminary results of designing and implementing a novel runtime architecture that, when fully mature, will support AI-out-HPC hybrid applications. We design and prototype our architecture in the context of the Low-Dose Understanding, Cellular Insights, and Molecular Discoveries (LUCID) research project~\cite{engel2024evaluating,umeike2025scaling} to study the effects of low-dose radiation on human cells to understand their underlying mechanisms and develop new therapeutic strategies. Ultimately, LUCID will deliver hybrid workflows for discovering novel cancer therapeutics, using large language models (LLMs) on exascale platforms to develop large-scale models trained on the vast low-dose radiation literature, automatic generation of experimental hypotheses, models, and pipelines for image classification, and hybrid workflows with surrogates and simulations of potential therapeutics.

Consistent with the growing interest in many research domains of integrating ML into scientific workflows~\cite{chen2024position}, we begin by focusing our design on general-purpose abstractions and a runtime architecture to support hybrid workflows. While training novel LLMs still requires ad-hoc and dedicated capabilities, there is a growing need for general-purpose capabilities to integrate diverse ML models in scientific workflows. That poses specific infrastructural and scaling challenges related to: (1) efficiently and effectively distributing inferences across multiple instances of possibly diverse models; (2) enabling each model instance to distribute inferences across multiple GPUs concurrently; (3) concurrently executing multiple model instances on specific portions of the available HPC resources; and (4) enabling the use of multiple model instances executing on local HPC and/or remote platforms.

Accordingly, we prototype runtime capabilities to scale hybrid workflows via service interfaces agnostic to the ML code they expose, instantiated on local or remote HPC/cloud platforms. We design and implement a runtime architecture based on proven abstractions (e.g., efficient scheduling of heterogeneous compute tasks and data staging) and novel, general-purpose service interfaces for local and remote execution of HPC/ML workflows at scale. We design an extensible, interoperable, and scalable architecture to add abstractions and capabilities as needed, integrate ML and HPC workflows implemented using different technologies and tools, and execute those workflows across local and remote sites.

This paper offers three main contributions: (1) an initial blueprint for a novel and performant runtime architecture; (2) a preliminary implementation of the architecture in the context of the LUCID project; and (3) an experimental evaluation of its weak and strong scalability on both local and remote deployment scenarios. Based on our present results, we have now started to implement three LUCID hybrid HPC/ML  pipelines, which will be used to evaluate the performance and scalability of our architecture. Once mature, our architecture will provide a foundation for designing and implementing general-purpose AI-out-HPC hybrid workflows and enable the efficient and effective integration of AI and HPC technologies in various scientific domains.

\section{Scenarios and Use Cases}\label{sec:uc}

LUCID requires a range of AI-out-HPC hybrid workflows, from those specifically designed to train large-scale LLMs on exascale HPC machines to classifying extensive collections of scientific papers, as well as more traditional ML pipelines and hybrid workflows that connect HPC tasks with either local or remote ML tasks. Here, we detail three pipelines that guide the design and implementation of our HPC/ML runtime architecture. While these pipelines have different requirements, they can all benefit from task-level parallelization and/or synchronicity, achievable through general-purpose service interfaces that enable efficient resource utilization at scale. These pipelines' data and/or compute requirements justify using HPC resources and runtime capabilities to integrate traditional HPC workloads with concurrent training and/or inference tasks.

\subsection{Cell Painting pipeline}\label{uc:cp}

This pipeline classifies radiation dose levels by analyzing cell-painting microscopy images using a fine-tuned ViT (Vision Transformer) model. It includes two stages: data processing and model training for signature detection.

The first stage processes and augments a cell-painting dataset ($\sim$1.6 TB) containing images that capture morphological changes in cells exposed to various radiation levels. Given the dataset's size, storage and transfer are managed using the Globus service. This stage doesn't require intensive computing power to normalize the raw images and apply augmentations such as rotation, cropping, flipping, and contrast adjustments. Data augmentation can be performed online to eliminate the need to store additional image data and allow for more diverse, randomly augmented data during training.

The second stage uses processed data for fine-tuning a ViT model, pre-trained on a large-scale natural image dataset (e.g., ImageNet). This model identifies key morphological signatures correlating with different radiation dose levels, transforming raw images into structured feature representations. This training phase involves GPU-intensive computations, thus starting only when sufficient processed data are available. The training is iterative, driven by hyperparameter optimization using the Optuna~\cite{akiba2019optuna} framework. This stage conducts multiple training iterations, exploring various hyperparameter configurations (e.g., learning rate, batch size, weight decay, and dropout rate).

To maximize efficiency, the pipeline is designed to run asynchronously and concurrently. Data preparation (including preprocessing and feature extraction) and model training (including hyperparameter optimization and fine-tuning) operate asynchronously while multiple models are trained concurrently, optimizing hyperparameters. Implementing each stage as a service enables asynchronous communication through dedicated APIs and concurrently instantiates various hyperparameters and training processes. This modular, service-based approach allows for dynamic resource allocation and release, ensuring efficient resource utilization.

\subsection{Signature detection pipeline}\label{uc:sd}

This three-stage pipeline analyzes DNA variants from 15 samples (each $\sim$300 MB VCF files) exposed to low-dose ionizing radiation. Its purpose is to identify radiation-induced mutational patterns and potential pathways of biological significance. By combining variant annotations with pathway databases and dose-response data, the pipeline provides insight into how low-dose radiation influences gene regulation and molecular functions.

In the first stage, the pipeline invokes the \texttt{Ensembl Variant Effect Predictor} (VEP)~\cite{EnsemblVEP} to annotate each sample’s VCF data. A single VEP run for one sample takes 1--5 minutes and requires $\sim$3 GB of memory. VEP can be run locally or via a REST interface on local and/or remote resources. Local VEP invocations are independent and can be concurrently executed at runtime. Exposing VEP via a service API enables asynchronous execution of the first and second stages.

In the second stage, annotated variants are combined with known pathways (e.g., KEGG~\cite{kanehisa2002kegg} and/or GO~\cite{chen2017prediction}) to identify significantly enriched genes, pathways, or molecular functions. This step relies on Python (e.g., \texttt{pandas}, \texttt{numpy}, and \texttt{scipy}) modules and is typically CPU-intensive. Runtime per sample remains in the order of minutes but can be parallelized across multiple cores.

In the third stage, additional tasks integrate the above results with temporal/dose information, producing dose-response insights. The resulting intermediate outputs (typically CSV files in the kilobyte to megabyte range) capture dose-response associations and serve as inputs for further analyses and visualization. Each additional analysis and visualization can run asynchronously and scale up as needed.

Looking ahead, the pipeline will incorporate LLMs to mine relevant literature (e.g., 10,000 curated papers, $\sim$20 GB) and generate hypotheses about gene regulation and putative biological signatures of low-dose radiation. This ML-based step will require scalable, on-demand runtime services for inference and knowledge extraction, adding a mixed workload of CPU- and GPU-intensive tasks to the pipeline.
As with use case~\ref{uc:cp}, Signature Detection will benefit from a concurrent and asynchronous execution and communication model. Service-based approaches will enable such asynchronicity, ensuring more efficient resource utilization.

\subsection{Uncertainty quantification (UQ) pipeline}\label{uc:uq}

This pipeline systematically evaluates the uncertainty in LLM inferences, comparing performance among different models and UQ methods. It consists of three stages: data preparation, multiple fine-tuning-based UQ tasks, and post-processing. In the first stage, existing datasets are preprocessed to be fed into the data pipeline of each UQ sub-task. Currently, the dataset contains approximately 3.4 MB of plain text formatted as question-and-answer pairs, but larger datasets are planned. This stage needs only negligible computing power compared with the next stage.

In the second stage, we perform a series of fine-tuning–based UQ tasks organized into a three-level hierarchy. At the innermost level, we evaluate various UQ methods (e.g., Bayesian LoRA, LoRA ensemble), benchmarking their performance. The middle level includes multiple random seeds for each UQ method, allowing for robust statistical analysis. Finally, at the outermost level, we compare the performance of different large language models (LLMs) such as Llama and Mistral. Performance-wise, each level should execute with maximal task concurrency while load balancing across UQ methods.
Each task utilizes $\sim$5–-60 GB of GPU memory, depending on the specific model size and LoRA configuration.

In the third stage, results from the second stage are aggregated and used to compute some metrics to summarize the performance of different UQ methods and models. This is another computational cheap stage.

\begin{table*}[htbp]
  \centering
  \caption{Use cases. Pipeline descriptions, resource requirements, and service-based implementation.}
  \label{tab:usecases}
  \footnotesize
  \begin{tabularx}{0.95\textwidth}{@{}>{\raggedright\arraybackslash}p{0.4cm}
                               >{\raggedright\arraybackslash}p{4cm}
                               >{\raggedright\arraybackslash}X
                               >{\raggedright\arraybackslash}p{2cm}
                               >{\raggedright\arraybackslash}p{3cm}@{}}
    \toprule
    \textbf{ID} &
    \textbf{Pipeline Name} &
    \textbf{Stage Name} &
    \textbf{Resource Type} &
    \textbf{Enable as Service} \\
    \midrule
    \multirow{2}{*}{\textbf{1}}                     &
    \multirow{2}{*}{Cell Painting}                  &
    Data pre-processing \& augmentation             &
    CPU                                             &
    Yes                                             \\
                                                    &
                                                    &
    Model training with hyperparameter optimization &
    GPU                                             &
    Yes                                             \\
    \midrule
    \multirow{3}{*}{\textbf{2}}                     &
    \multirow{3}{*}{Signature Detection}            &
    Data Preparation                                &
    CPU                                             &
    Yes                                             \\
                                                    &
                                                    &
    Mutation Detection Analysis                     &
    CPU                                             &
    No                                              \\
                                                    &
                                                    &
    LLM-based signature comparison                  &
    GPU                                             &
    Yes                                             \\
    \midrule
    \multirow{3}{*}{\textbf{3}}                     &
    \multirow{3}{*}{Uncertainty Quantification}     &
    Data Preparation                                &
    CPU                                             &
    Yes                                             \\
                                                    &
                                                    &
    UQ methods with three-level parallelism         &
    GPU                                             &
    No                                              \\
                                                    &
                                                    &
    Post-processing                                 &
    GPU                                             &
    Yes                                             \\
    \bottomrule
  \end{tabularx}
\end{table*}

\section{Design and Implementation}\label{sec:arch}

Our architecture and its underlying HPC/ML coupling runtime capabilities are motivated by the learning everywhere paradigm~\cite{fox2019taxonomy} and designed based on the building block engineering approach~\cite{turilli2019middleware}. As such, we enable various HPC/ML modalities and software integration at runtime, middleware, and workflow levels.

\begin{figure}
  \centering
  \includegraphics[trim=0 3 0 1,clip,width=0.49\textwidth]{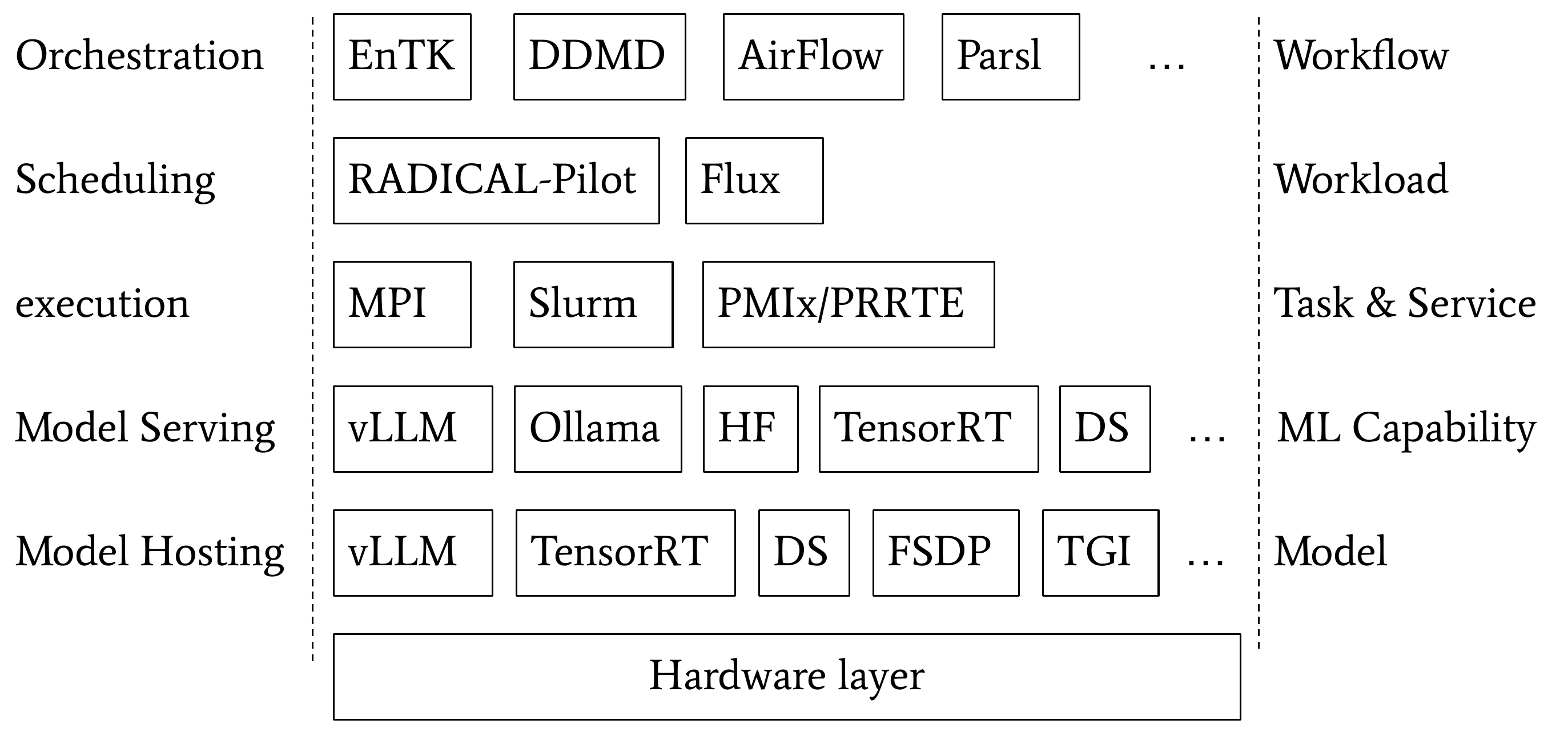}
    \caption{HPC/ML capabilities, technologies, and entities stack. Each layer contributes distinct capabilities to manage entities that enable scalable, concurrent execution of heterogeneous HPC/ML workflows. We display only a representative subset of the technology ecosystem available (HF = Hugging Face Transformers and DS = DeepSpeed).}\label{fig:stack}
\end{figure}

Fig.~\ref{fig:stack} describes the HPC/ML design space, offering a layered representation of capabilities (left column), a sample of the technologies that implement those capabilities (center boxes), and the logical entities on which those capabilities operate (right column). We design an architecture that enables the end-to-end stacking of all the needed capabilities. Crucially, we integrate and extend existing technologies, expanding the current scientific computing ecosystem, promoting reusability, and avoiding wasteful effort duplication.

While we are designing an architecture that, in principle, will satisfy the functional, nonfunctional, and performance requirements of the learning everywhere paradigm and its growing list of motifs~\cite{jha2022ai,brewer2024ai}, in this paper, we prototype only a subset of the required capabilities. Specifically, we focus on the requirements of the LUCID use cases presented in \S\ref{sec:uc}, supporting the execution of general-purpose service tasks that expose arbitrary ML capabilities. In that way, we can implement the ML stages of a pipeline (e.g., use case \S\ref{uc:cp}) and enable distributed inference across local and remote model instances, agnostic towards the type of model and of inference/prompt request (e.g., \S\ref{uc:sd}). Importantly, we maintain HPC workflow/workload orchestration capabilities, supporting the need for task concurrent execution in our use cases (e.g., \S\ref{uc:sd} and \ref{uc:uq}).

We assume workflow or pipeline applications are described via workflow management systems (Fig.~\ref{fig:stack}, orchestration and workflow). We then develop our runtime HPC/ML capabilities within RADICAL-Pilot~\cite{merzky2021design}, a pilot-enabled~\cite{turilli2018comprehensive} runtime system designed to manage distributed, heterogeneous and dynamic workloads in HPC environments at up to exascale~\cite{titov2024scaling} (Fig.~\ref{fig:stack}, execution and model serving). RADICAL-Pilot integrates with several workflow systems (e.g., AirFlow, EnTK, Parsl, or BV-BRC) to obtain an end-to-end solution for implementing our use cases.

RADICAL-Pilot provides robust scheduling and execution capabilities but lacks general-purpose model serving or hosting capabilities. We extend RADICAL-Pilot by introducing service-oriented runtime abstractions, enabling an arbitrary number of services, each exposing a unified API for ML models. Each service operates independently of the technology used to serve and host a specific model. We allow specification of which resources each model should utilize and whether these resources should be dynamically shared among models. It is important to note that while extending RADICAL-Pilot, we maintain full backward compatibility. Users can take advantage of the new features directly within RADICAL-Pilot, with no need for additional software systems.

While we design our service interface to integrate with many existing model serving and hosting technologies, in this paper, we use Ollama, avoiding the complexities of alternatives that would enable efficient parallelization on HPC (e.g., vLLM, TensorRT,  or DeepSpeed). This is consistent with the need first to develop and characterize a general-purpose service application programming interface (API) designed to scale in local and remote scenarios. We will then extend our capabilities by evaluating and characterizing the integration and performance of more complex technologies. Ultimately, we will complete the implementation of our architecture by extending our service capabilities to support distributed online model training (e.g., PyTorch FSDP).

RADICAL-Pilot operates with tasks as units of work, executed independently of each other and following a stateful paradigm. Implementation of the service infrastructure includes extending RADICAL-Pilot's \textit{Task} abstraction into \textit{Service Task} with corresponding service management and interface capabilities. That allows for managing services as regular tasks while connecting them to dedicated communication channels to receive control commands. New capabilities allow controlling whether a service is launched before or after other services and/or computing tasks and monitoring each service's availability at runtime. For example, in workflows, services often have to be started before any computing task and run throughout the workflow's duration~\cite{yokelson2024enabling}.

Services require specific functionalities. Each service should expose a well-defined interface (e.g., a REST API) to tasks (i.e., clients) and be available to receive client calls anytime. As such, services require readiness and liveness management. In response to these requirements, we implement a \textit{Service Base Class} in RADICAL-Pilot and use the ZeroMQ communication infrastructure to enable API calls between services and clients. When implementing the LUCID use cases, we will create a new class, exposing methods for ML model handling via a general-purpose API. We will then scale the number of services by launching, monitoring, and terminating each service instance as a task within RADICAL-Pilot.

\begin{figure}
  \centering
  \includegraphics[trim=0 3 0 1,clip,width=0.49\textwidth]{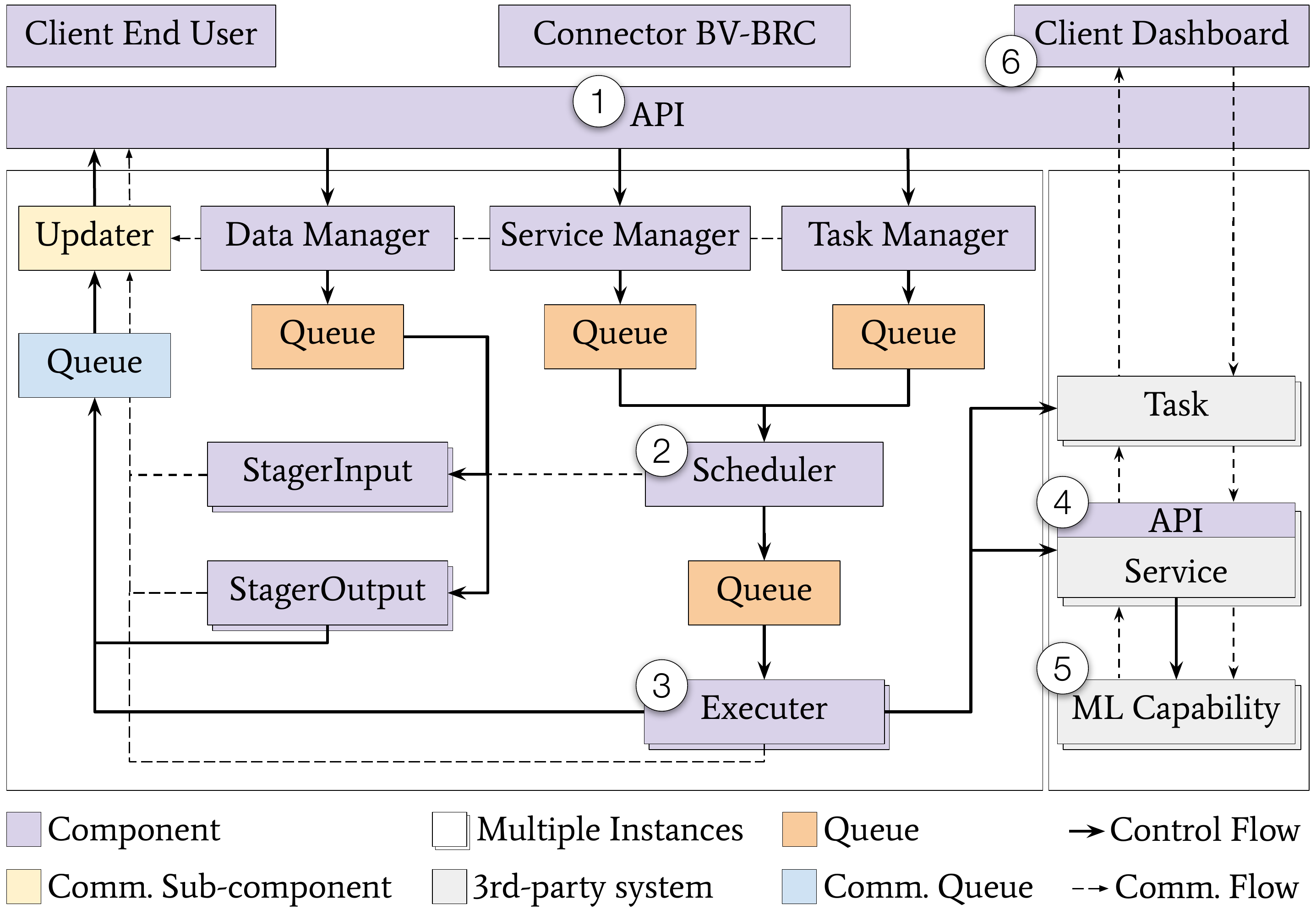}
    \caption{Runtime architecture to support HPC/ML coupling. We extended RADICAL-Pilot with service-specific capabilities to enable large-scale deployment of ML capabilities on HPC. Numbers indicate the execution model of the service capabilities enabled by this architecture.}\label{fig:arch}
\end{figure}

Fig.~\ref{fig:arch} shows how we extended RADICAL-Pilot architecture~\cite{merzky2021design} with service-specific components. We implemented a \T{ServiceManager} with the capabilities described above, complementing the existing \T{TaskManager}, and collected existing data capabilities into a \T{DataManager}. Service, task, and data managers can access properties and state information about the entities they manage. In this way, they can derive readiness relations and guarantee that data are staged and each service is running and available to receive client (i.e., computing tasks) requests. We extended the existing \T{Scheduler} to enact priority relations between services and tasks and leveraged the existing \T{Executor} capabilities.

RADICAL-Pilot's execution model now enables users to submit \T{ServiceDescription} and \T{TaskDescription} via a unified API (\Cref{fig:arch}~\ccircled{1}), requesting their placement via the \T{Scheduler} (\Cref{fig:arch}~\ccircled{2}) and execution via the \T{Executor} (\Cref{fig:arch}~\ccircled{3}). Each service exposes a specific API (\Cref{fig:arch}~\ccircled{4}) and instantiates specific capabilities, e.g., an ML model (\Cref{fig:arch}~\ccircled{5}). Users (or third-party middleware components) get information about services and tasks via dedicated communication channels (\Cref{fig:arch}~\ccircled{6}).

Our architecture is agnostic regarding the type of service, its API, and the capabilities it exposes. RADICAL-Pilot enables the definition of an arbitrary number of services, the specification of whether they operate locally or remotely, the allocation of each service to specific local resources (e.g., within a single node or across multiple nodes), and the determination of how long each service instance should be available and when it should be terminated. After characterizing the performance of our architecture, we will implement high-performance services, replacing Ollama with model serving and hosting technologies specifically designed for HPC, and suitable for prototyping the LUCID use cases.

\section{Performance Characterization}\label{sec:exp}

We characterize the performance of the HPC/ML capabilities described in \S\ref{sec:arch}. We use three metrics: (1) \textbf{Bootstrap Time (BT)}: the time taken by a certain number of services to become available (BT); (2) \textbf{Response Time (RT)}: the time taken by each service to acknowledge an inference request; and (3) \textbf{Inference Time (IT)}: the time taken by a model to serve an inference request. BT and RT characterize the overheads of our implementation because they measure time spent on the infrastructure and not on computing the scientific payload. IT measures the time taken by a model exposed via a service interface to serve an inference to a task that sends a prompt to that model via its service interface. We measure BT, RT, and IT in seconds and consider their distribution across multiple task, service, and model instances. This allows us to measure their averages and observe outliers and long tails.

We consider local and remote deployment scenarios and parameterize our experiments accordingly. In the local scenario, we acquire HPC resources via a pilot job and then use those resources to launch several service instances and a set of tasks that use those services. Each service exposes an arbitrary ML model, and one or more tasks send inference requests to that model via the service interface. In this scenario, we tune six experimental parameters to measure the strong and weak scaling of our HPC/ML infrastructure: the total amount of available resources (i.e., the size of the pilot), the number of services and tasks that can concurrently execute on those resources, the number of inferences that each task can request, and the number of concurrent inference requests and of concurrent inferences each service can receive and perform.

In the remote scenario, we make no assumptions about how the services and models are instantiated. This is consistent with the design approach described in \S\ref{sec:arch} and the use cases described in \S\ref{sec:uc}. As such, our experiments with remote service capabilities do not measure BT but only RT and IT, scaling the number of concurrent remote services available and the number of concurrent inference requests and concurrent inferences each service can receive and perform.

\begin{table*}
    \caption{Experiment Setup. Experiment 1 measures bootstrap overheads (BT) while experiment 2 and 3 measure local and remote service response time (RT) and inference time (IT), respectively.}\label{tab:experiments}
	\centering
    \footnotesize
	\begin{tabularx}{0.93\textwidth}{llllllllll}
	\toprule
    \textbf{ID}                &  
	\textbf{HPC Platform}      &  
    \textbf{Task Type}         &  
    \textbf{Model}             &  
    \textbf{Model Deployment}  &  
    \textbf{\#Tasks}           &  
    \textbf{\#Models}          &  
    \textbf{\#Cores/Pilot}     &  
    \textbf{\#GPUs/Pilot}      &  
    \textbf{Scaling}           \\ 
	\midrule
	\textbf{1}   &  
    Frontier     &  
    n/a          &  
    llama 8b     &  
    local        &  
    n/a          &  
    1--640       &  
    640          &  
    40           &  
    weak         \\
    %
	\multirow{2}{*}{\textbf{2}}  &  
    Delta                       &  
    NOOP                        &  
    noop                        &  
    local                       &  
    1--16                       &  
    1--16                       &  
    256                         &  
    16                          &  
    strong/weak                 \\
                                &
    Delta and R3                &  
    NOOP                        &  
    noop                        &  
    remote                      &  
    1--16                       &  
    1--16                       &  
    256                         &  
    16                          &  
    strong/weak                 \\
    %
	\multirow{2}{*}{\textbf{3}}  &  
    Delta                       &  
    inference                   &  
    llama 8b                    &  
    local                       &  
    1--16                       &  
    1--16                       &  
    256                         &  
    16                          &  
    strong/weak                 \\
                                &
    Delta and R3                &  
    inference                   &  
    llama 8b                    &  
    remote                      &  
    1--16                       &  
    1--16                       &  
    256                         &  
    16                          &  
    strong/weak                 \\
	\bottomrule
	\end{tabularx}
    \UP
\end{table*}

\subsection{Experiments Design and Parameterization}

We design three experiments (see Table~\ref{tab:experiments}) to measure the functional and performance capabilities of the runtime prototype presented in \S\ref{fig:arch}. Experiment 1 measures the startup overhead required to make a growing number of model instances available to an HPC workflow. That overhead has multiple time components needed to: (1) start the service that hosts the model instance; (2) load the model into memory and initialize the model; and (3) communicate the service endpoints to the task. We collectively define the sum of 1--3 as the bootstrap time (BT).  BT depends on the number of concurrent model instances being started. Note that we measure BT only for local ephemeral models. Remote models are usually persistent on dedicated resources and do not need to be bootstrapped for application instances.

Experiment 2 measures the time a growing number of concurrent local and remote model instances take to acknowledge an inference request, i.e., response time (RT). We investigate RT by implementing a NOOP model, which will immediately reply without performing any actual inference. As with BT, RT also has several time components required for: (1) a request to communicate from task to service instance; (2) the service to parse that request; (3) the service to process the request and form a reply; and (4) the task to receive the service's reply. Communication time usually contributes the most significant part of RT, and we investigate its dependency on network latency and its scaling as a function of the number of concurrent requests per instance.

Finally, Experiment 3 measures the time to serve an increasing number of inference requests by concurrent local and remote model instances (IT). We measure IT for each model instance and report the average and distribution of the inference time across all instances. This allows us to investigate the scaling behavior of running multiple concurrent model instances. Currently, services are single-threaded, and, as such, they only handle one request at a time, queuing further incoming requests. We will drop this simplification when investigating more scalable ML serving and hosting technologies (see Fig.~\ref{fig:stack}), enabling experiments that measure the scalability of specific model serving and hosting capabilities in terms of concurrency within and across multiple compute nodes.

We execute our experiments on three platforms: OLCF Frontier, NCSA Delta, and R3, a cloud-based server on which we expose ML capabilities via REST and ZeroMQ interfaces. OLCF Frontier enables scaling experiment 1 to 640 concurrent service instances to measure the impact of concurrent service instance instantiation. Delta enables local and local/remote scenarios, with both NOOP and actual ML capabilities to measure service and ML model response time.

\subsection{Experiment 1: Scaling of Local Service Bootstrap Time}

In this experiment, we launch several service instances, each using one GPU on Frontier OLCF.  We increase the number of instances during each experiment run (1, 2, 4, 8, 20, 40, 80, 160, 320, and 640 instances. For each run, we measure the individual overhead contributions for launching the service executables on their target resources (\T{launch}, orange), loading and initializing the LLM models (\T{init}, green), and publishing the service endpoints (\T{publish}, red).

\begin{figure}
  \centering
  \includegraphics[trim=0 3 0 1,clip,width=0.45\textwidth]{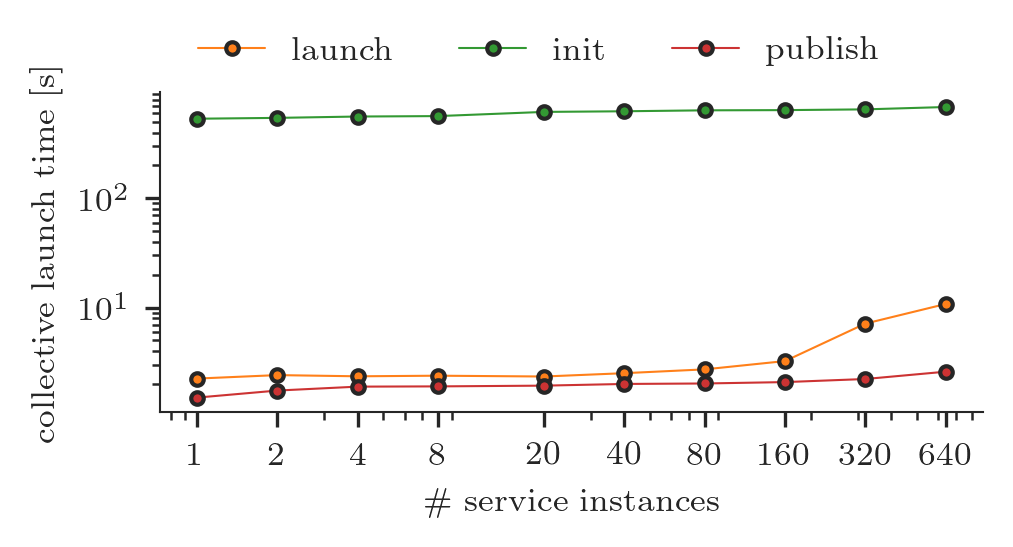}
    \caption{Service Bootstrap Times. Individual contributions to the overall bootstrap time for an increasing number of local service instances.}\label{fig:service-launch}
\end{figure}

Figure~\ref{fig:service-launch} shows that the time to launch service instances on their target resources (\T{launch}) is almost constant, up to 160 instances. Beyond that, an increasing system-level launch overhead is observed. Our preliminary analysis suggests that the observed increase is due to MPI startup time, but targeted experiments are needed to confirm this. In all cases, the time used to publish the service endpoints (\T{publish}) remains smaller than the launch time. Both communication and launching times are negligible compared to the time needed to load and initialize the model instances (\T{init}). We used \texttt{ollama} to host and serve the \texttt{llama-8b} model and did not invesitgate it's startup overhead. As already stressed in \S\ref{sec:arch}, we remain agnostic towards the service and hosting capabilities, enabling the execution of alternatives to \texttt{ollama} and \texttt{llama-8b}.

Currently, LUCID use cases require fewer than 640 concurrent services; thus, the observed overheads will not impact their implementation. If future use cases require further scaling of concurrent services, we will utilize both resource partitioning and asynchronous execution, as has already been successfully done. in~\cite{merzky2022raptor,titov2022radical}.

\subsection{Experiment 2: Strong and Weak Scaling of Local and Remote Service Response Time (RT)}

Experiment 2 examines the effect of network latencies on service response time. No actual inferences are made. Instead, the service instances execute a \T{noop} function that returns immediately. This way, each service immediately replies to incoming service requests with a static response. The experiment quantifies the individual contributions to the overall response time, including the time to communicate the request to the service and return the response to the client (\T{communication}, orange), the time the service requires to queue, deserialize and parse the request, and to serialize the reply message (\T{service}, green), and the time to execute \T{noop} and to form the reply message (\T{inference}, red).

\begin{figure}
  \centering
  \includegraphics[trim=0 3 0 1,clip,width=0.45\textwidth]{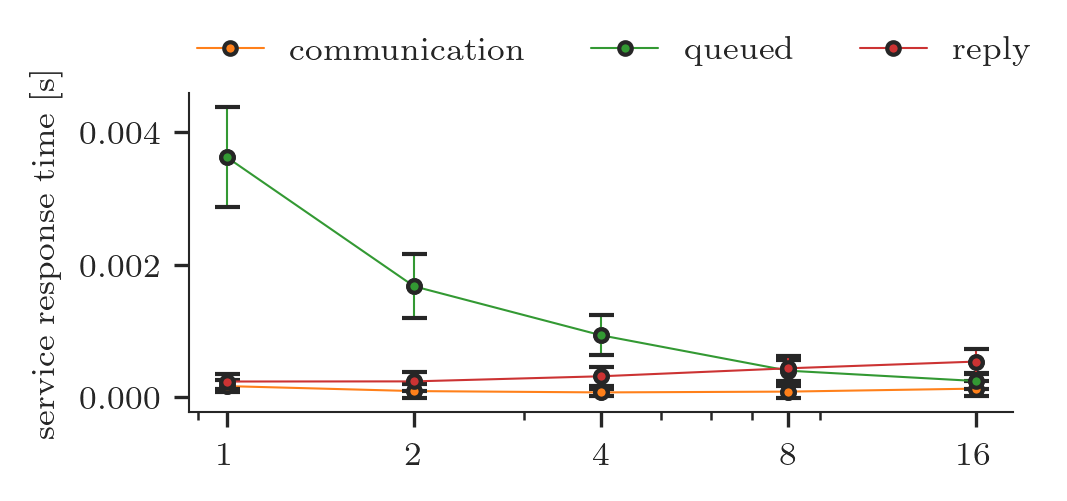}
  \includegraphics[trim=0 3 0 1,clip,width=0.45\textwidth]{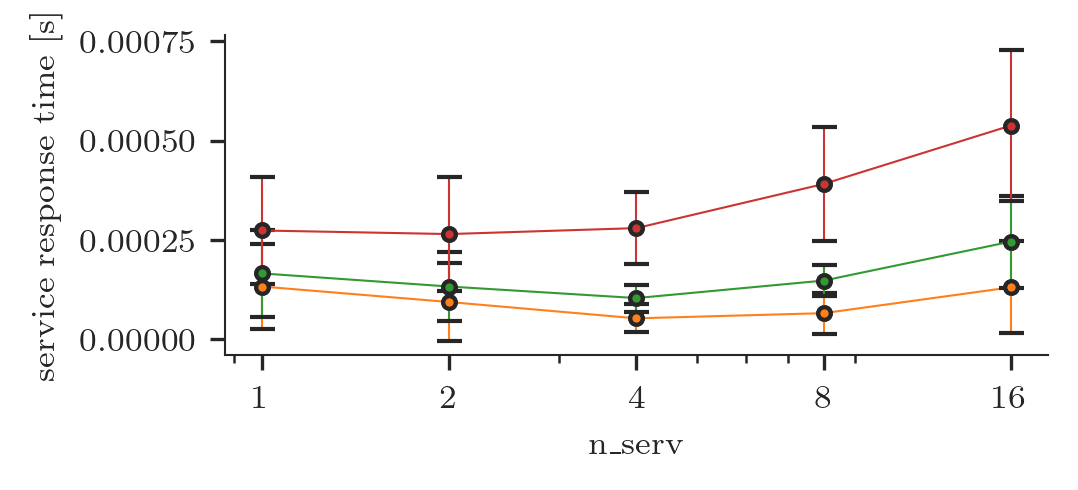}
  \caption{Service Response Times for \textit{local} NOOP inference calls. Strong scaling (top, number of clients == 16) and weak scaling (bottom, number of services == number of clients).}\label{fig:local_noop}
\end{figure}

\begin{figure}
  \centering
  \includegraphics[trim=0 3 0 1,clip,width=0.45\textwidth]{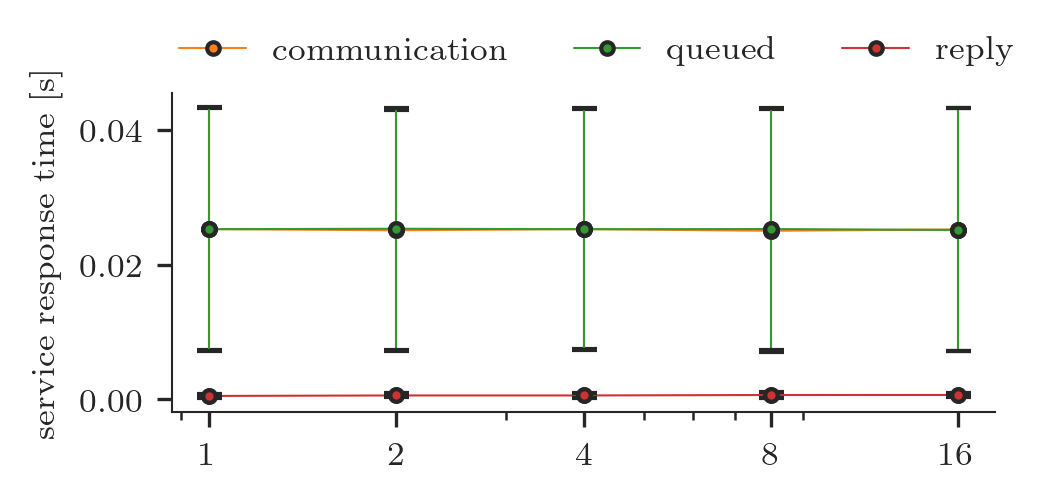}
  \includegraphics[trim=0 3 0 1,clip,width=0.45\textwidth]{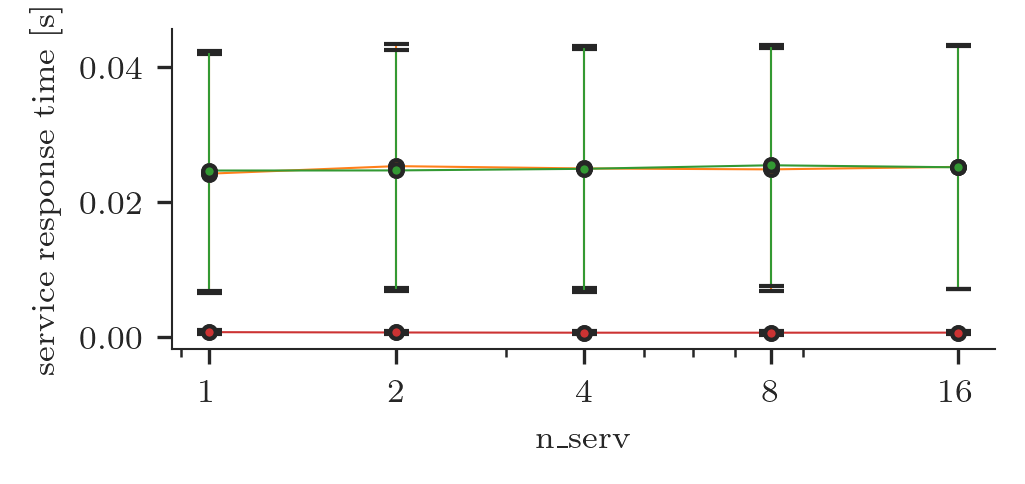}
  \caption{Service Response Time for \textit{remote} NOOP inference calls. Strong scaling (top, number of clients == 16) and weak scaling (bottom, number of services == number of clients).}\label{fig:remote_noop}
\end{figure}

We investigate local services, running on the same platform as the client tasks (inter-node-latency: 0.063 ms +/- 0.014 ms), and remote services, running on a different platform---i.e., R3---(node-to-node-latency: 0.47 ms +/- 0.04 ms).  As shown in~\Cref{fig:local_noop,fig:remote_noop}, we vary the number of service instances and client tasks to measure the strong and weak scaling of the service response time, with each client sending a fixed number of inference requests (1024). For strong scaling, we used 16/1, 16/2, 16/4, 16/8, and 16/16 clients/services, thus using a constant number of requests and increasing the number of services. For weak scaling, we used 1/1, 2/2, 4/4, 8/8, and 16/16 clients/services, thus increasing the overall load (number of requests) with the number of services.

\Cref{fig:local_noop,fig:remote_noop} confirm that the contributions to the overall response time are negligible compared to both remote and local network latency. In a scenario of continuous and distributed inference requests, the impact of latencies can be reduced by increasing the number of concurrent service instances, which effectively raises the number of potential requests in flight simultaneously over the network.

\up\subsection{Experiment 3: Strong and weak Scaling of Local and Remote Model Inference Time (IT)}

This experiment measures the configuration we will use in our use cases: a number of client applications submit multiple inference calls to a set of local or remote service instances. We again measure the contributions to the service response time---sending the inference request and receiving the response (\T{communication}, orange), the time the request is handled and queued by the service (\T{service}, green)---but we also measure the time the service takes to use its backend (LLM Meta Llama 3 with 8B parameters) to generate responses (\T{inference}, red).  Otherwise, the experiment setup is identical to that of experiment 2.

\begin{figure}
  \centering
  \includegraphics[trim=0 3 0 1,clip,width=0.45\textwidth]{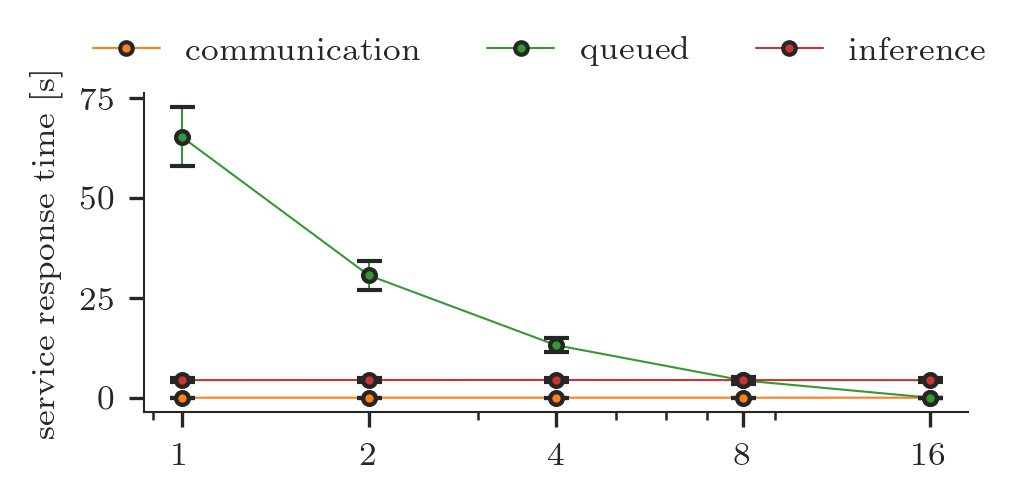}
  \includegraphics[trim=0 3 0 1,clip,width=0.45\textwidth]{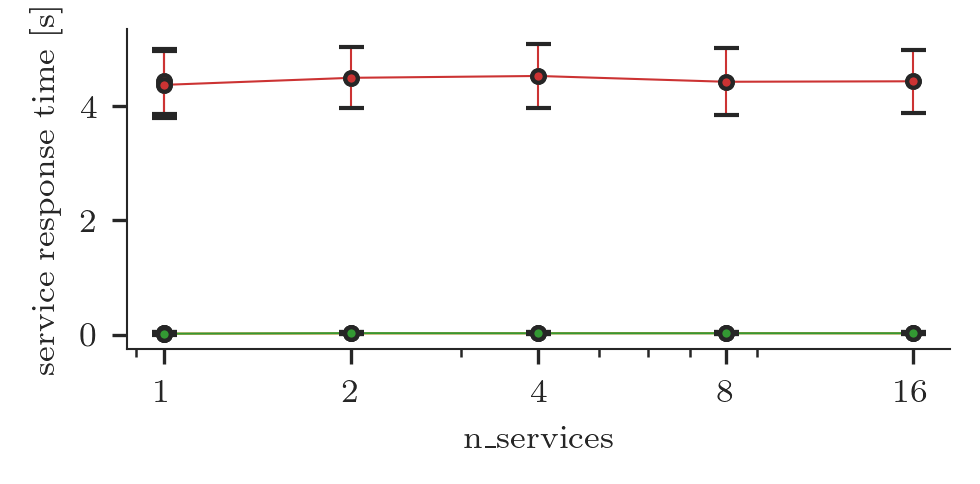}
    \caption{Service Response Times for \textit{remote} LLAMA inference calls. Strong scaling (number of clients == 16) and weak scaling (number of services == number of clients).}\label{fig:remote_llama}
\end{figure}

While network latency dominates the overall response time in experiment 2, in experiment 3, \jhanote{error: should we remove experiment 2, or should we remove experiment 3?} the inference time is significantly larger than all other time contributions (see Fig.~\ref{fig:remote_llama}, bottom, weak scaling). This implies that model locality is a secondary concern for actual inference calls regarding overall response time. Potential optimizations to model performance, and consequently pure inference times, are more essential. The strong scaling plot (top) shows that the service queues client requests because, as expected, the backend is too slow.

\subsection{Future Directions}

The experimental results show that our runtime transparently supports both local and remote ML service instances with negligible overhead, similar to what was measured with RADICAL-Pilot before the implementation of service capabilities~\cite{merzky2021design}, and at the scale required by the driving use cases. We also discussed how, should future scaling requirements arise, we can further iterate our design to reduce those overheads through resource partitioning and asynchronous execution. Relatively long inference durations may lead to a backlog of requests if the number of service instances is insufficient. Although remote and local service instances exhibit different response times, that difference becomes negligible when the actual inference times dominate the overall response time.

Our experiments' goal is not to measure the performance of our use cases. Instead, before implementing our use cases, we must ensure that our architecture is viable and that its base performance is adequate to satisfy the requirements described in \S\ref{sec:uc}. Future experiments will require expanding on the current model serving and hosting capabilities. So far, we only used Llama 8b (hosted by Ollama) as an LLM instance. While they guarantee a straightforward way to serve and host a very capable model, they are not designed to efficiently and effectively support our target use cases at scale. Further, we do not optimize model concurrency or utilize request queuing or any other latency hiding, and we employ only a rudimentary load balancing. Finally, so far, we limited the number of concurrent clients since the limited service concurrency would not provide any further insight.

We consistently plan to expand our experiments to include more specialized and high-performing capabilities like those listed in Fig.~\ref{fig:stack}. This will allow us to optimize model serving and hosting, enhancing service-level request concurrency, inference speed, and overall service throughput. Expanding our experiments with a growing number of clients will also necessitate better load balancing across all available service instances, dynamically rerouting requests to less used service instances. With these capabilities, we will prototype our use cases and measure their scalability by varying the duration, type, and number of concurrent HPC tasks. We will compare the results of these experiments to the HPC-only baseline performance published in~\cite{merzky2021design,turilli2019characterizing,titov2022radical}.

\section{Related Work}\label{sec:related}

Several software solutions exist to couple HPC and ML on supercomputers at middleware and runtime levels. DeepHyper~\cite{balaprakash2018deephyper} focuses on ML model training, hyperparameter optimization, and neural architecture search in HPC environments. While DeepHyper provides optimization capabilities, it does not address the challenges of runtime inference scaling and HPC/ML workflow integration. Our work introduces a runtime service model that supports dynamic ML capabilities scaling, ensuring efficient execution of ML models across diverse computing resources.

Parsl offers fine-grained GPU-multiplexing capabilities exposed via the Function as a Service paradigm~\cite{dhakal2023fine}. Currently, our architecture focuses on executable tasks, i.e., self-contained processes placed on specific HPC nodes and containing the logic to access one or more CPUs/GPUs. As such, our runtime prototype and Parsl are complementary, serving two different use cases. As we already integrate RADICAL-Pilot and Parsl~\cite{alsaadi2022radical}, we will extend our architecture to support both executable and function tasks~\cite{merzky2022raptor}.

Colmena~\cite{ward2025employing} is a Python library that adds a layer to conventional workflow systems to enable intelligent orchestration of ML-driven simulations and experiments in HPC environments. Colmena enables adaptive task scheduling and real-time inference integration but focuses primarily on steering computational campaigns using active learning. Our work introduces runtime service capabilities that generalize HPC/ML coupling, allowing scalable distributed ML serving and hosting rather than workflow steering.

Ray Serve~\cite{schroder2024comparison,moritz2018ray} provides a scalable model-serving framework optimized for distributed inference workloads. Ray Serve is well-suited for ML applications that require efficient request distribution and dynamic scaling. However, it primarily targets cloud instead of supercomputers and lacks tight integration with HPC resource management mechanisms. Our approach addresses this gap by enabling ML serving and hosting services to execute efficiently on HPC clusters while utilizing remote ML capabilities, possibly instantiated on cloud platforms. In that way, we ensure interoperability and scalability across different execution environments.

RedisAI~\cite{boyer2022scalable,brewer2021production} is an inference-serving framework that integrates AI workloads into in-memory databases, optimizing inference latency for real-time applications. While RedisAI efficiently handles model storage and execution within Redis, it does not support distributed inference scaling across multiple HPC nodes. Our architecture enables multi-node inference scaling, allowing ML models to serve requests efficiently across local and remote computing environments, while avoiding the bottleneck of single-threaded communication across multiple RedisAI instances.

\section{Conclusions}\label{sec:conclusion}

In this paper, we design, prototype, and evaluate a scalable runtime architecture to support AI-out-HPC workflows. We focus on integrating ML capabilities with traditional HPC workloads. By extending RADICAL-Pilot with service-oriented abstractions, our system enables the scalable and efficient execution of hybrid HPC/ML workflows across local and remote HPC/cloud platforms.

Through experimental evaluation, we demonstrated the performance of our architecture in three key aspects: (1) bootstrap overheads for initializing local ML capabilities; (2) response times for ML inference requests; and (3) inference execution scalability. Our results confirm that service-based execution enables asynchronous and concurrent resource utilization, ensuring efficient workload distribution. Moreover, while network latencies impact remote inference, overall execution time is dominated by model inference duration rather than communication overhead, suggesting optimization efforts should prioritize inference efficiency.

Looking ahead, we will further develop our runtime capabilities following the blueprint architecture presented in \S\ref{sec:arch} and the insight gained with our experiments. Specifically, we will integrate ML serving and model hosting capabilities by integrating HPC-specific/compatible technologies such as vLLM, TensorRT, and DeepSpeed, improving concurrency and inference throughput. Additionally, future work will explore adaptive resource scheduling to dynamically load-balance HPC and ML workloads across heterogeneous computing environments. Finally, we will scale our experiments based on the prototyping of the use cases presented in \S\ref{sec:uc}, measuring scalability across a varying number of client tasks, service instances, model parallelism, and concurrent request rates.


\noindent {\footnotesize {{\bf  Data and analysis} scripts at:
{\footnotesize \url{https://github.com/radical-experiments/lucid}}}}

\noindent {\footnotesize {{\bf Acknowledgements} US DOE DE-AC02-06CH11357 (LUCID), NSF-2103986 and 1931512.}

\small

\up\bibliographystyle{IEEEtran}
\bibliography{lucid}

\end{document}